\documentclass[preprint,aip,apl,letterpaper]{revtex4}
\pdfoutput=1

\usepackage{amssymb}
\usepackage{amsmath}
\usepackage{bbm}
\usepackage{graphicx}
\usepackage{subfigure}
\usepackage{makeidx}
\usepackage{braket}
\usepackage{nth}

\usepackage[OT2,T1,T2A]{fontenc}
\DeclareSymbolFont{cyrletters}{OT2}{wncyr}{m}{n}
\DeclareMathSymbol{\Sha}{\mathalpha}{cyrletters}{"58}

\newcommand{\vecb}[1]{\mbox{\boldmath$#1$}}
\newcommand{\hatb}[1]{\mbox{\boldmath$\hat{#1}$}}

\newcommand{\wcsq}{\left(\frac{\omega}{c}\right)^2}

\newcommand{\besselj}[2]{J_{#1}(#2)}

\DeclareMathOperator{\jinc}{jinc}

\newcommand{\rr}{\vecb{r}}
\newcommand{\rrp}{\vecb{r'}}

\newcommand{\kk}{\vecb{k}}

\newcommand{\fingerprintingSchematics}{L1106847R1.pdf}
\newcommand{\figIScale}{0.95}

\newcommand{\figFingerprintingCurvesA}{L1106847R2.pdf}
\newcommand{\figIIScale}{0.95}

\newcommand{\eeta}{\vecb{\eta}}
\newcommand{\qq}{\vecb{q}}

\newcommand{\rrpp}{\vecb{r''}}

\newcommand{\hankel}{H^{(1)}_0}

\newcommand{\egrid}{\delta\epsilon_\text{gr}}

\newcommand{\etarget}{\delta\epsilon_\text{t}}

\begin{document}

\title{Super-resolution Spatial Frequency Differentiation of Nanoscale
  Particles with a Vibrating Nanograting}


\author{Leonid Alekseyev, Evgenii Narimanov}
\affiliation{Electrical and Computer Engineering Department, Purdue University, West Lafayette, IN 47907}
\email{evgenii@purdue.edu}
\author{Jacob Khurgin}
\affiliation{Electrical and Computer Engineering Department, Johns Hopkins
University, Baltimore, MD 21218}



\begin{abstract}
  We propose a scheme for detecting and differentiating deeply
  subwavelength particles based on their spatial features.  Our approach
  combines scattering from an ultrasonically modulated nanopatterend grating
  with heterodyne techniques to enable far-field detection of high spatial
  frequency Fourier components.  Our system is sensitive to spatial features
  commensurate in size to the patterning scale of the grating.  We solve the
  scattering problem in Born approximation and illustrate the dependence of
  the signal amplitude at modulation frequency on grating period, which
  allows to differentiate between model nanoparticles of size $\lambda/20$.
\end{abstract}

\maketitle 

Detection and differentiation of micro- and nanoscale particles is of great
practical importance in medical and biological research, clinical
diagnostics, and many other areas.  Historically, optical
methods have been the preferred strategy of investigating small objects,
since they allow rapid, cost-effective, noninvasive analysis.  As a result, a
multitude of imaging and detection modalities have been developed.  However,
many advantages of optical detection disappear when the objects under study
become substantially smaller than the wavelength, owing to the conventional
$\lambda/2$ diffraction limit.  This severely curtails the amount of
information that can be obtained about structure, function, and composition
of nanoscale particles by optical means.
%
%
%
%

In recent years, researchers have developed several approaches for optical
investigation of nanoscale particles below the diffraction limit. 
Broadly
speaking, those methods fall into one of two categories.  One set of
strategies aims to maximize the amount of information that can be inferred
about the subwavelength objects based on aspects of scattering that are not
affected by the diffraction limit.  For instance, particles can be detected
due to fluorescence or various other spectroscopic resonances~\cite{Hell2007}.  Other
approaches exploit the dependence of Rayleigh scattering intensity on
particle size~\cite{Daaboul2010,Ignatovich2006}.
The second category of methods involves broadening
the spatial frequency passband of the system~\cite{BrueckOE2007} by recovering
evanescent waves that ordinarily cannot propagate to the far-field detector.
For a number of years, this could only be achieved with optical scanning
probes~\cite{Dragnea2001,Keilmann1999,Planken2002}.   In the past decade,
intense research in nanophotonics and plasmonics led to approaches
towards subwavelength imaging that involve using metamaterials-based
devices, which promise to be low-cost and readily adaptable for a
variety of spectral regions.  A negative index ``superlens''~\cite{pendry}, for
instance, can transport the evanescent components of the
spatial spectrum across the metamaterial slab, while also transmitting the propagating waves.  However,
the superlens exhibits exponential sensitivity to losses, limiting
practical devices to near-field
operation~\cite{PodolskiyNarimanovNSSL}.  A more recent set of
experiments involved subwavelength scatterers placed in the near field
of a source.  By diffracting off the scatterers, evanescent waves
convert into propagating waves which could then be used to gather
information about the near-field spectrum~\cite{Durant2006} or to
attain subwavelength focusing~\cite{LeroseyFink2007}.  However, in
this far-field superlens~\cite{Durant2006}, the diffracted evanescent
waves mix with the existing propagating spectrum.  The resulting ambiguity leads to the loss
of information about the target.

In the present Letter, we propose an alternative approach to far-field
detection of nanoparticles, based on a device that converts evanescent waves
to propagating waves via scattering on an acoustically-modulated nanograting.
The salient feature of our approach is the fact that since the scattered
waves are also shifted in the frequency domain, they can be easily decoupled from the
existing propagating spectrum.  We describe the proposed system and calculate
its output signal to show that it can be used for differentiating subwavelength objects based on
their size, shape, and/or composition.

\begin{figure}
 \centerline{\scalebox{\figIScale}{\includegraphics{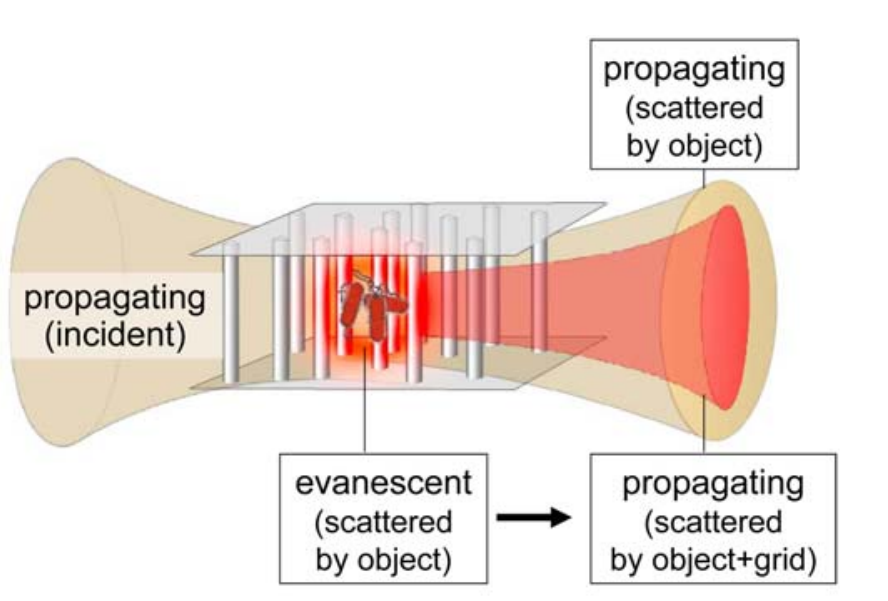}}}
 \caption{Schematics of the proposed system.  Evanescent components of
   radiation from the sample scatter from the nanostructure and propagate
   into the far field.  Vibration of the nanorods leads to a frequency shift
   which enables detection of these evanescent contributions using lock-in techniques.
} \label{fig:fingerprintingSchematics}
\end{figure}

The proposed detection system is shown in
Fig.~\ref{fig:fingerprintingSchematics}. The target object, embedded in a
patterned regular array of nanorods, is illuminated with a plane wave. The
spacing between the elements of the patterned array determines the resolution
of the system, as well as the maximum allowable size of the target object.
Because of this, the pictured device is particularly suitable for studying
severely subwavelength targets.  The nanorods are made to vibrate using a
high-frequency piezoelectric transducer.  Ultrasonic devices reaching 10~GHz
are within reach of contemporary technology~\cite{Lanz2005,Assouar2007},
making it possible to achieve frequency shifts exceeding the linewidth of a
narrowband optical source.

Due to patterning and utrasonic modulation, the dielectric function of the
nanorod array is periodic both in space and time. In the space
domain it can be represented as a Fourier series with large wave vectors ($q
\gg 2\pi/\lambda$).  Thus, components of the incident radiation that scatter
from subwavelength features of the target and are now evanescent (due to
acquiring large transverse wave vectors) can scatter again from the nanorod
array, reducing their transverse wave vector.  As such, these fields can
retain their propagating nature while carrying subwavelength information.  
Now, because of the periodic time dependence of the nanorod
permittivity (which we write as $\epsilon = \overline{\epsilon} + \delta
\epsilon(\rr-\eeta \cos\Omega t)$, with $\eeta$ being the amplitude and
$\Omega$ the frequency of the ultrasonic vibration), the frequency of these
scattered signals is shifted by $\Omega$, thereby separating them out from
the rest of the propagating waves and allowing their unique retrieval.

To analyze the proposed system, we will consider the Green's function
solution of time-harmonic Maxwell's equations.  For simplicity, we treat
a system that is translationally invariant in the direction of the
cylindrical nanorods, perfectly aligned along the $z$ axis.  The target
objects are also assumed to have infinite extent in the $z$ direction.
Further, we assume that the spatial dimensions and/or the index contrast of
the scattering system are small.  In this case, the solution to Maxwell's
equations can be represented by the Born series
\begin{equation}
  \label{eq:scatt_series}
  \begin{split}
    E(\rr) =  E_0(\rr) &+ \left(\frac{\omega}{c}\right)^2 \int G(\rr,\rrp)
    \delta\epsilon(\rr)  E_0(\rrp) d\rrp + \\
     &+ \left(\frac{\omega}{c}\right)^4 \int G(\rr,\rrp)
    \delta\epsilon(\rr) \int  G(\rrp,\rrpp)\delta\epsilon(\rrpp)E_0(\rrpp)
    d\rrpp d\rrp\\
    &+O\left((\delta\epsilon)^3\right),
    \end{split}
\end{equation}
where $E_0(\rr)=\exp(i \kk \cdot \rr)$ and $G(\rr,\rrp) \sim
H^{(1)}_0(k|\rr-\rrp|)$ is the usual Green's function for a 2D Helmholtz
equation, proportional to the Hankel function, and
$\delta\epsilon(\rr)=\epsilon(\rr)-\overline{\epsilon(\rr)}$ serves as a weak
scattering potential.  We retain the second order term in the Born series
because this is the lowest order in which the acoustically-shifted signal at
frequency $\omega+\Omega$ is in the propagation band of the system.
Separating the scattering potential into contributions from the grid and from
the target ($\delta \epsilon=\egrid+\etarget$), we can decompose
Eq.~\eqref{eq:scatt_series} into terms corresponding to the different
scattering processes:

\begin{equation}
  \label{eq:scatt_abbrev}
    E(\rr) \equiv E_0(\rr) - \frac{i}{4}\wcsq\left( I_\text{gr} +I_\text{t}\right)
    - \frac{1}{16}\left(I_\text{gr-gr} +I_\text{t-t} + I_\text{gr-t} +I_\text{t-gr} \right).
\end{equation}
Here, terms of the form $I_\text{x}$ and $I_\text{x-y}$ stand for Green's
function integrals expressing the contributions to the total field due to
single and double scattering.  In particular, 
\begin{equation}
  \label{eq:ib_explicit}
    I_\text{x-y}=\iint \delta\epsilon_\text{x}(\rrp)\delta\epsilon_\text{y}(\rrpp)\hankel(k|\rr-\rrp|)\hankel(k|\rrp-\rrpp|) \exp(i\kk\cdot\rrpp) d\rrpp d\rrp.
\end{equation}

The important terms in Eq.~\eqref{eq:scatt_abbrev} are $I_\text{gr-t}$ and
$I_\text{t-gr}$.  The fields corresponding to those terms carry
information about the high spatial frequency components of the target objects
that have been downshifted into the propagation band of the system and,
furthermore, have been ``tagged'' with the modulation frequency of the grid.
Interference between these shifted signals at frequency $\omega \pm \Omega$
with the illuminating wave allows to retrieve the information through lock-in
detection (provided the modulation frequency $\Omega$ is used as a reference).
When we put in an explicit expression for a vibrating
grid, both terms lead to the same expression:

\begin{equation}
  \label{eq:i-gr-t}
I_\text{gr-t} \sim  \frac{e^{ikr}}{\sqrt{kr}}
 \sum_{m=-\infty}^\infty e^{i m \Omega}\left[\sideset{}{'}\sum_{\qq_n} i^m \besselj{m}{\qq_n\cdot\vecb{\eta}}
 \jinc\left(\frac{aq_n}{2\pi}\right) \frac{\etarget(\kk_\text{obs}-\kk-\qq_n)}{q_n^2-2\kk_\text{obs}\cdot\qq_n}\right].
\end{equation}
Here, the following assumptions have been made: the grid is composed of a
regular 2D array of cylindrical rods with radius $a$ and spacing
$\vecb{\Lambda}$ ($a < \Lambda \ll \lambda$); $\qq_n\sim 1/\vecb{\Lambda}$
are the grid reciprocal lattice vectors, $\jinc(r) \equiv J_1(2\pi
  r)/\pi r$, $\kk_\text{obs} \equiv k\hatb{r}$ is a propagation vector
aligned with the detector, and $m$ is the diffraction order.  Since the
detector is assumed to be in the far field, Hankel asymptotics were used.

\begin{figure*}
 \centerline{\scalebox{\figIIScale}{\includegraphics{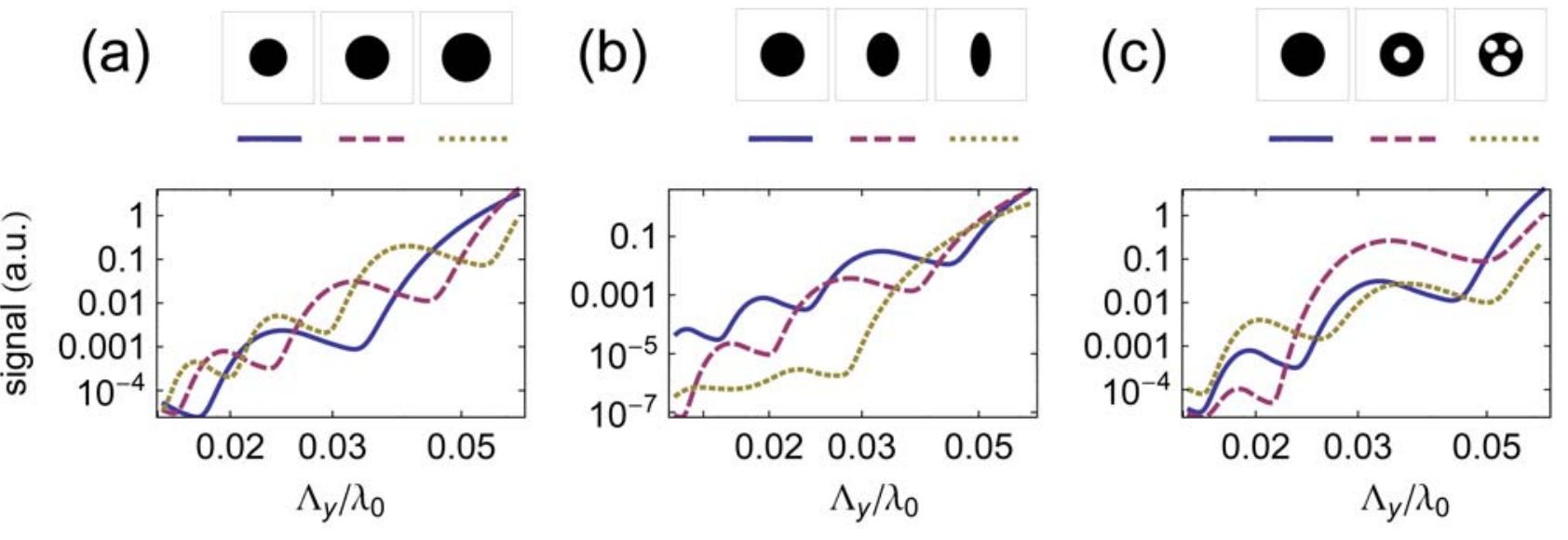}}}
 \caption{Simulated signal as a function of grating period in the $y$
   direction as size, shape, and structure [panels (a), (b), and (c)
   respectively] of the model target (cylinder with $d=\lambda/20$) are
   altered. Different line types (solid, medium dashing, fine dashing)
   correspond to the different target cross sections, as illustrated. }
 \label{fig:FingerprintingCurvesA}
\end{figure*}

In addition, we expect a realistic sample to contain many target objects
randomly distributed in the nanorod grid and randomly oriented.  As a result,
the Fourier representation of the dielectric function $\etarget(\qq)$ will
contain a sum over the phase factors corresponding to optical path length
difference between the scatterers.  Because the relative positions of the
scatterers are random, these phase factors form a mean zero distribution.
Due to the finite integration time at the detector, as well as its finite
aperture, the average contribution from the random phases can be shown to
approach zero~\cite{AlekseyevThesis}.  Taking $m=1$ in Eq.~\eqref{eq:i-gr-t} and squaring to compute
the intensity signal, we obtain the following expression for the detected
signal at frequency $\Omega$:

\begin{equation}
\label{eq:an_mod_squared_averaged_expandJ1}
  |A|^2 \propto N \sideset{}{'}\sum_{\qq} \left|\qq\cdot\vecb{\eta}
 \jinc\left(\frac{aq}{2\pi}\right)
 \frac{\etarget^\text{av}(q)}{q^2}\right|^2  + O\left(\frac{N}{\sqrt{T}}\right),
\end{equation}
where $N$ is the number of target objects, $T$ is the integration time, and
$\etarget^\text{av}(q)$ is the angular average (due to random orientations)
of the spacial frequency representation of an individual target.

To illustrate the ability of the proposed system to distinguish between
subwavelength particles, in Fig.~\ref{fig:FingerprintingCurvesA} we plot the
signal computed from Eq.~\eqref{eq:an_mod_squared_averaged_expandJ1} as a
function of the grating period in the $y$ direction ($\Lambda_y$).  In order
to achieve effective tunability of this period in a physical device, it could
be manufactured as a chirped grid in which the local period varies slowly on the
scale of the wavelength.  (Within the effective scattering volume, the local
period should be approximately constant; any imperfections in the grid will
decrease device performance.)  Note that it would be sufficient to vary the period
along one direction only; in the other direction the nanorod spacing could be
arbitrary.  In particular, it could be made large enough to allow objects to
be translated in the device, e.g. via microfluidics.  

As a basic model target, we take a cylinder with
diameter $\lambda/20$, well below the diffraction limit.  We then explore how
the computed signal differs as we perturb the target's size, shape, and as we
introduce some inner structure to the particle [panels (a), (b), and (c) of
Fig.~\ref{fig:FingerprintingCurvesA}].  The abscissas indicate the local
period of the nanostructure in the $y$ direction in units of $1/\lambda$.
The
curves, plotted on the log-log scale, illustrate the differences between
signals for the different conformations of our model particle, depicted above
the panels.  The secular trend of the curves indicates a power
law decrease of the signal as spatial frequencies increase.  

The clear differences between the curves in each panel of
Fig.~\ref{fig:FingerprintingCurvesA} suggest that changes in either the size,
shape, or structure of the targets on a deeply subwavelength scale can be
detected with our proposed approach.  We note, however, that due to angular
averaging over multiple randomly oriented targets, this scheme is sensitive
only to the average radial spatial frequency distributions in the sample,
which are not necessarily unique.  Thus, these signals do not provide
unambiguous signatures for arbitrary targets.  However, even with this
limitation it should be possible to extract useful information given some
prior knowledge about the sample under study.  For instance, given a mixture
containing the differently-sized particles of
Fig.~\ref{fig:FingerprintingCurvesA}(a), we could provide an estimate of the
size of the smallest component in the mixture by observing the location of
the first minimum of the signal as the period $\Lambda_y$ decreases.  Similarly, for a
circle/ellipsoid family of Fig.~\ref{fig:FingerprintingCurvesA}(b), we could
estimate the eccentricity of the ``narrowest'' ellipsoid.

If it is known that only \textit{one} out of a set of possible targets is
present, its identity could be revealed by comparing the data to known
calibration signals.  

Finally, we can also imagine other experiments, in which the shape or structure of the target
is changing in real time under the influence of some external mechanism (e.g.\ mitochondria
undergoing calcium overload~\cite{Boustany2002} or cells undergoing
apoptosis~\cite{Mulvey2007}).  Looking at the curves of
Fig.~\ref{fig:FingerprintingCurvesA}(b) it's easy to see that as the
ellipsoidal targets become deformed into spheres (a good model for the
calcium-induced alterations in mitochondrial morphology~\cite{Boustany2002}), the
signal changes rather dramatically.  Thus, the proposed system could be used
to detect real-time target morphology changes on a deeply subwavelength scale.

In conclusion, we have proposed a system that enables differentiation of
nanoscale particles by their sub-diffraction-limited spatial spectrum
components, which can be detected in the far field by utilizing scattering
from an ultrasonically modulated grating.  We note that the model system
treated here is one out of many possible configurations, and that many
potential implementations exist, including layered planar systems.
Compared to existing techniques, our scheme allows to create a much richer spatial
frequency fingerprint for nanoscale objects.


\end{document}